\documentclass[prx,letterpaper,aps, nobalancelastpage,twocolumn,superscriptaddress,nofootinbib,10pt]{revtex4-2}
\usepackage{color,braket,graphicx,amsmath,amssymb}
\usepackage{times}
\usepackage{amsfonts}
\usepackage[caption=false]{subfig}
\usepackage{ragged2e}
\usepackage{framed}
\usepackage{tikz}
\usepackage{dsfont}
\usepackage[inline]{enumitem}
\usepackage{amsthm}
\usepackage{color,xcolor}
\usepackage[normalem]{ulem}

\newcommand{\be}{\begin{equation}}
\newcommand{\ee}{\end{equation}}

\usepackage{orcidlink}
\usepackage{tikz}
\usetikzlibrary{shapes.geometric, arrows}
\tikzstyle{startstop} = [rectangle, rounded corners, minimum width=1.5cm, minimum height=0.8 cm,text centered, draw=black, fill=red!40]
\tikzstyle{io} = [trapezium, trapezium left angle=70, trapezium right angle=110, minimum width=0.6cm, minimum height=0.6cm, text centered, draw=black, fill=blue!30]
\tikzstyle{process} = [rectangle, minimum width=3cm, minimum height=1cm, text centered, draw=black, fill=orange!30]
\tikzstyle{decision} = [diamond, minimum width=3cm, minimum height=1cm, text centered, draw=black, fill=green!30]
\tikzstyle{arrow} = [thick,->,>=stealth]

\begin{document}
\title{Autonomous Optical Alignment of Satellite-Based Entanglement Sources using Reinforcement Learning}
\author{Andrzej Gajewski \orcidlink{0000-0001-7248-2561}} 
\email{andrzej.rafal.gajewski@gmail.com}  
\noaffiliation
\author{Robert Oku\l a \orcidlink{0000-0002-9941-6991}}
\email{rbrt.okula@gmail.com} 
\affiliation{Department of Algorithms and System Modeling, Faculty of Electronics, Telecommunications and Informatics, Gda\'{n}sk University of Technology, Gabriela Narutowicza 11/12, 80-233 Gdańsk, Poland}
\affiliation{Department of Physics, Stockholm University, 106 91 Stockholm, Sweden}  
\author{Marcin Paw\l owski \orcidlink{0000-0002-8611-947X}}
\email{marcin.pawlowski@ug.edu.pl}
\affiliation{International Centre for Theory of Quantum Technologies, University of Gdańsk, Jana Bażyńskiego 1A, 80-309 Gdańsk, Poland} 
\author{Akshata Shenoy H. \orcidlink{0000-0002-6703-8383}}
\email{akshata.shenoy@ug.edu.pl}
\affiliation{International Centre for Theory of Quantum Technologies, University of Gdańsk, Jana Bażyńskiego 1A, 80-309 Gdańsk, Poland}   

\begin{abstract}
Quantum entanglement distributed via satellites enable global-scale quantum communication. However, onboard sources are susceptible to misalignment due to dynamical orbital conditions. Here, we present two recalibration techniques for efficient generation of high quality entanglement using a periodically poled lithium niobate (PPLN)-based spontaneous parametric down-conversion (SPDC) source with minimum intervention. The first is a heuristic algorithm (HA) which mimics the manual alignment process in a laboratory. The second is based on reinforcement learning (RL). Our simulation demonstrates superior performance of RL with AUC=0.9119 compared to HA's 0.7042 in the  modified ROC analysis (60 min threshold). RL achieves perfect alignment in 10 min as opposed to HA's 30 min. Both the methods operate within feasible satellite constraints, offering scalable automation for complex quantum communication scenarios.
\end{abstract}

\maketitle

\section{Introduction}
Satellite Quantum communication employed in tandem with fiber-optic networks \cite{LKB21}, paves the way for scalable communication connecting even the farthest corners of the Earth \cite{SJG+21,B13}. Conventional fiber-optic networks suffer from significant transmission losses beyond certain distances \cite{AEE+23}. Hence, without quantum repeaters \cite{muralidharan2016optimal} to preserve quantum signals over long ranges \cite{frtv-p5zy}, such networks become impractical for large-scale applications \cite{kucera2024demonstration}, changing the underlying security assumptions \cite{doi:10.3233/JCS-2010-0373, Xiao2025, Gaidash:22}. In contrast, free-space quantum communication based on a satellite link experiences comparatively lower losses \cite{SB19}. When utilized alongside terrestrial fiber-optics\cite{chen2021integrated}, satellites constitute a powerful resource for global quantum communication. Thus, enabling unconditional security from the fundamental principles of quantum mechanics \cite{lu2022micius,P21}.

Such ideas have been demonstrated experimentally through the exchange of quantum signals between Low Earth Orbit (LEO) satellites and ground stations \cite{BMH+13}, using both polarization \cite{VBD+15,TCF+17} and time-bin \cite{VDT+16} encoding schemes for quantum protocols. A major milestone was achieved with the distribution of entanglement over distances exceeding \(1200\,\mathrm{km}\) via satellite--ground links \cite{YCL+17}, as well as the realization of an intercontinental quantum-secure communication link between China and Austria \cite{LYL+17}. More recently, microsatellite platforms have demonstrated real-time quantum key distribution with multiple ground stations, highlighting the practical nature of space-based quantum systems \cite{Li2025microsat}. The realization of global quantum networks with enhanced operational capabilities will require efficient generation, distribution and routing of quantum states between satellite and ground nodes \cite{PAL+05,Vinet2025reconfigurable,DeSantis2025parallel}.

In satellite-based entanglement distribution \cite{MI20}, photon pairs generated onboard, are transmitted to ground stations through free-space, which are inherently subject to atmospheric effects. They include weather-dependent losses, beam distortions and elevated background noise during daylight operation \cite{VBD+15}. Apart from atmospheric propagation, the space environment itself introduces additional challenges \cite{joarder2025entanglement}. An orbiting satellite alternates between illuminated and shadowed regions of the Earth, experiencing rapid thermal changes and mechanical perturbations \cite{badas2024opto} that can significantly impact stability of the source. The performance of SPDC sources are highly sensitive to such environmental fluctuations \cite{bedington2016nanosatellite}. Furthermore, variations in temperature modifies the refractive index of the nonlinear crystal, modifying the phase-matching conditions. This directly affects the brightness and entanglement quality \cite{lasota2020optimal} of the source. Simultaneously, thermally induced stress and vibration can lead to misalignment of critical optical elements, such as interferometric components, further degrading the source. Consequently, for sustained high-quality entanglement generation in satellite orbits, active compensation for both the crystal and optical component parameters is needed.

Calibration is defined as the process of systematic adjustment of parameters of a device to ensure its optimal operation. It is therefore an indispensible step in experimental physics. However, conventional calibration procedures are typically manual, time-consuming and impractical for autonomous operations \cite{frank2017autonomous}, especially in dynamic environments. With increasing complexity and number of controllable degrees of freedom in experiments \cite{saha2025automating}, scalable and automated calibration is desirable. In this work, we model and characterize a PPLN–based entanglement source designed for space applications \cite{YCL+17}. We present post-launch recalibration methods to maintain efficient SPDC with minimum intervention under realistic conditions. First, HA inspired by standard manual alignment procedures is presented. Building on this, a RL algorithm demonstrating improved performance over HA is shown. This is quantified using receiver operating characteristic (ROC) curves and the corresponding area under the curve (AUC).

The rest of this paper is organized as follows. Results and discussion are presented in Sections~\ref{results} and~\ref{discuss} respectively. The algorithms are described in the Methods section. In particular, Subsection~\ref{exp_set-up} details the optical setup used to realize the entanglement source. The HA employed for active optimization to generate highly correlated photon pairs is described in Subsection~\ref{handalgo}, while Subsection~\ref{ralgo} outlines the training and implementation of the RL algorithm. Additional details, metric used and hyperparameter specifications are provided as Supplementary Notes~1--3.

\section{Results}
\label{results}

To quantitatively compare the performance of HA and RL, we have formulated a performance metric that captures the final calibration accuracy at a fixed adjustment window of 60 min and its speed of convergence to the correct position.

\begin{figure}[htbp]
    \centering

    \textbf{a}\quad Modified ROC for the human-interaction-based algorithm\\
    \includegraphics[width=0.6\linewidth]{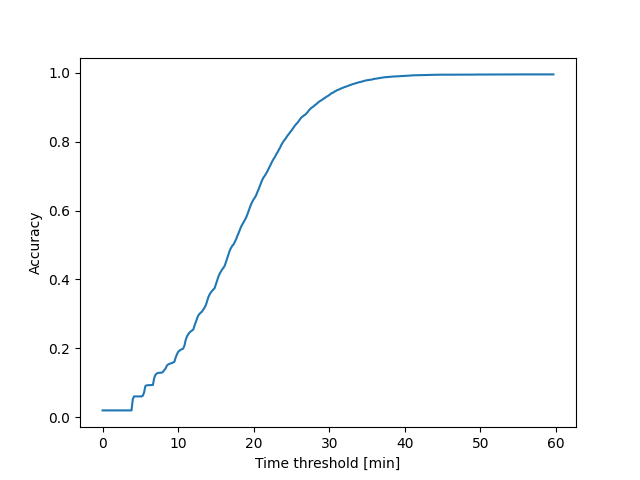}

    \vspace{1em}

   \textbf{b}\quad Modified ROC for the best-performing RL agent\\
   \includegraphics[width=0.6\linewidth]{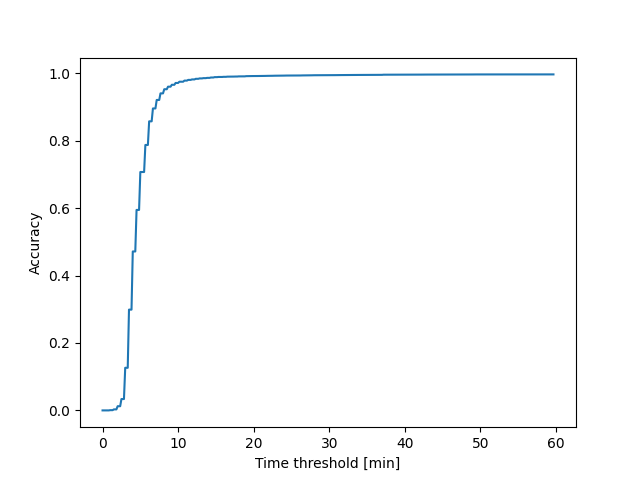}

   \caption{Modified ROC curves showing the accuracy metric defined in Eq.~\eqref{eq:A} as a function of calibration time.}
    \label{fig:ROC}
\end{figure}

We have defined a modified Area Under the Curve (AUC)~\cite{BRADLEY19971145} based on a modified receiver-operating characteristic (ROC) curve, relating the calibration time threshold $t^{max}_i$ to the accuracy quantifier
\begin{equation}
\label{eq:A}
A = \frac{N_{t_j < t^{max}_i}}{N_{all}},
\end{equation}
where $N_{all}$ denotes the total number of generated starting points and $N_{t_j < t^{max}_i}$ is the number of cases for which alignment is completed in time less than the threshold $t^{max}_i$. The metric attains a value of $A = 1.0$ when all the trials converge within the specified time limit. Though the maximal threshold is fixed to be 60 min, a faster convergence results in higher AUC values. The RL-based approach achieves a higher AUC score ($AUC_{RL}^{max} = 0.9119$) compared to the HA ($AUC_{HA}^{max} = 0.7042$) as shown in Fig.~\ref{fig:ROC}. The RL agent reaches a near perfect alignment within approximately 10 min, whereas the HA requires around 30 min.

\section{Discussion}
\label{discuss}

We have presented two complementary techniques for maintaining efficient entangled photon pair generation in satellite-based sources subject to misalignment under orbital conditions. The HA mimicking manual alignment serves as a baseline for the RL algorithm. It is seen that, RL demonstrates superior adaptability with more efficient source parameter optimization and faster convergence. This superior performance of RL can be attributed to its efficient exploration–vs-exploitation and faster policy stabilization (Fig.~\ref{fig:ROC}). Hence, resulting in an increase in mean reward and decrease in episode length. The modified AUC metric used proves particularly valuable in capturing the temporal efficiency of these methods, that a single fixed-threshold accuracy measure would otherwise miss. Importantly, both approaches operate within experimentally feasible design constraints and remain compatible with established quantum communication protocols. These results establish RL as a robust and practical alternative for automated alignment in realistic satellite quantum communication scenarios.

\section{Methods}
\subsection{The entanglement source}
\label{exp_set-up}

Quantum entanglement is generated via SPDC, a nonlinear optical process in which a pump photon with angular frequency $\omega_p$ and wave vector $\vec{k}_p$ annihilates inside a nonlinear crystal producing lower-energy signal $(\omega_s, \vec{k}_s)$ and idler $(\omega_i, \vec{k}_i)$ photons. A PPLN crystal with large second-order nonlinear coefficient ($\sim$25 pm V$^{-1}$) enabling bright, tunable and high-quality entangled photon-pair generation \cite{steinlechner2012high} is chosen.

Efficient type-0 SPDC obeys the quasi-phase-matching condition
\begin{eqnarray}
\omega_p &=& \omega_s + \omega_i, \nonumber \\
0 &=& \vec{k}_p - \vec{k}_s - \vec{k}_i \pm \vec{K},
\label{eq:qpm}
\end{eqnarray}
where, $\vec{K} = 2\pi/\Lambda$ is the grating vector for poling period $\Lambda$. Successful phase matching produces an entangled biphoton wavefunction~\cite{KWB+09}.

We simulate a PPLN crystal with $\Lambda=19.388~\mu$m, $10$~mm length, pumped at $775$~nm to produce collinear $1550$~nm signal and idler photons. Temperature dependence simulation reveals a sharp reduction in SPDC below $25^{\circ}$C (Supplementary Fig.~(\ref{fig:SPDC})) showing phase-mismatch sensitivity. See Supplementary Note~1 for details.

\subsection{Heuristic alignment algorithm}
\label{handalgo}
\begin{figure}[htbp]
    \centering
    \includegraphics[width=1.0\linewidth]{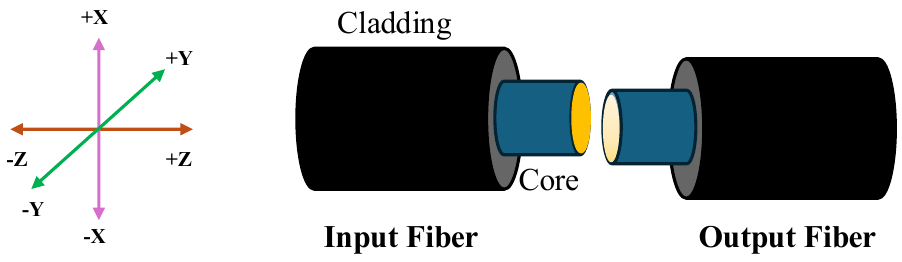}
    \caption{End-to-end optical coupling between two fibers. The input fiber is adjusted axially ($Z$) and transversely ($XY$) relative to a fixed output fiber to maximize coupling efficiency.}
    \label{fig:schematic}
\end{figure}

In the PPLN-based entanglement source, a Sagnac interferometer is usually employed to generate stable polarization entanglement. Fiber-optic couplers in the interferometer are highly sensitive to misalignment during satellite transit. We implement a HA that automates alignment, maximizing coupling efficiency by optimizing the axial $Z$ and radial $XY$ degrees of the input fiber relative to a fixed output fiber (Fig.~\ref{fig:schematic}).

The HA mimics standard manual laboratory procedures of starting from a pre-defined optimal $Z$ position. Initiating a blind deterministic jump to the theoretical optimal $Z_0=1580\mu$m along the $Z-$axis, followed by an initial measurement (upto $120$s integration time) to verify a signal strength above the background threshold noise. Next, alterating searches and refinements along the $Z-$axis and the radial $XY$ plane analogous to a four-point square pattern is implemented. The improvements are accepted when a standardized metric $W$ exceeds the statistical thresholds associated with maximum mean photon count rates or minimum time constraints, corresponding to $99.5\%$ (axial) or $99.9\%$ (radial) confidence levels. For final convergence, a theoretical maximum of $90\%$ is set for the counts within a time period of $3600$s. For details on control logic, decision threshold and the metric $W$, see Supplementary Note ~2.

\subsection{Reinforcement learning}
\label{ralgo}

In RL, an agent interacts with an environment by choosing actions that maximizes a cumulative reward \cite{sutton_reinforcement_2018}. Unlike supervised learning, RL does not require labelled input--output pairs. The agent learns a strategy through direct interactions with the environment, guided solely by a reward as feedback.

RL problems are Markov decision processes (MDPs) defined by a tuple $(S, A, p, r, \gamma)$ where $S$ is the state space, $A$ is the action space, $p$ is the transition dynamics, $r$ is the reward function and $\gamma$ is a discount factor. At each time step, the agent observes the current state $s \in S$ of the environment, selects an action $a \in A$ through a policy $\pi$, while transitioning to a new state $s'$ and receiving a reward $r(s,a,s')$. The objective of the learning process is to find an optimal policy maximizing the overall expected cumulative reward \cite{10.1561/2200000071}.

In this work, we utilize a Soft Actor--Critic (SAC) algorithm~\cite{haarnoja_soft_2018}, an off-policy actor--critic method, that jointly optimizes the expected reward and the policy entropy for efficient exploration and stable learning. The actor proposes actions observing the state of the environment, while the critic network evaluates these actions by estimating their expected return. SAC employs temporal-difference learning \cite{Sutton1988} with bootstrapping for data-efficiency and online learning. SAC's ability to adapt to variations and noise in the environment provides a flexible and scalable alternative to manual optimization.

\subsection{MDP formulation}
We now formulate the alignment problem of two fiber-optic couplers as a MDP.

\paragraph{State space}
The state space in the MDP framework comprises of positions of the input fiber in cylindrical coordinates $(r(t), \theta(t), z(t))$, corresponding to Cartesian coordinates $(x(t), y(t), z(t)) = (r \cos \theta, r \sin \theta, z)$ and a count rate $c(t)$, measured over a time step $t_{\mathrm{step}}$ (hyperparameter),
\begin{equation}
    s(t) = \bigl(r(t), \theta(t), z(t), c(t)\bigr),
\end{equation}
with $c(t) = c\bigl(r\cos\theta(t), r\sin\theta(t), z(t)\bigr)$.

\paragraph{Action space}
Actions are continuous input fiber displacements in the cylindrical coordinates. Normalized actions are output by the policy,
\begin{equation}
    a(t) = \bigl(\Delta r_{\mathrm{norm}}(t),\, \Delta \theta_{\mathrm{norm}}(t),\, \Delta z_{\mathrm{norm}}(t)\bigr) \in [-1, 1]^3,
\end{equation}
scaled linearly by maximum step sizes, yielding physical displacements,
\begin{align}
    \Delta r(t)      &= \Delta r_{\mathrm{norm}}(t)\, r_{\mathrm{step}}^{\max}, \\
    \Delta \theta(t) &= \Delta \theta_{\mathrm{norm}}(t)\, \theta_{\mathrm{step}}^{\max}, \quad \theta_{\mathrm{step}}^{\max} = \pi, \\
    \Delta z(t)      &= \Delta z_{\mathrm{norm}}(t)\, z_{\mathrm{step}}^{\max}.
\end{align}
The position updates as
\begin{align}
    r(t+1)     &= r(t) + \Delta r(t), \\
    \theta(t+1)&= \theta(t) + \Delta \theta(t), \\
    z(t+1)     &= z(t) + \Delta z(t),
\end{align}
with optional clipping $r(t+1) \leftarrow \max(r(t+1), 0)$ preventing negative radii. This symmetric scaling permits bidirectional movement along all axes.

\paragraph{Reward function}
Reward guides the agent towards configurations with higher photon counts. At each time step, the environment returns
\begin{equation}
    r(t) = b(t) - p,
\end{equation}
where $p>0$ is a small step penalty and $b(t) \ge 0$ a bonus for count-rate gain. The instantaneous count rate is
\begin{equation}
    c(t) = \frac{n_{\mathrm{count}}(t)}{\Delta t}, \quad 
    \Delta c(t) = c(t) - c(t-1).
\end{equation}

To reward substantial improvements, a threshold
\begin{equation}
    s_b = \frac{c_{\max}}{l_{\mathrm{bonus}}},
\end{equation}
partitions the expected maximum count $c_{\max}$ into $l_{\mathrm{bonus}}$ equal segments. The bonus is then
\begin{equation}
    b(t) = b_0 \cdot \max\Bigl(0, \bigl\lfloor \Delta c(t)/s_b \bigr\rfloor\Bigr),
\end{equation}
where $b_0$ (hyperparameter) sets the reward per crossed threshold. This promotes decisive moves while penalizing inefficient ones.

\subsection*{Optimization and training}
Superior performance of RL algorithms require careful parameter optimization and neural network training, particularly in complex environments. This ensures stability, exploration-vs-exploitation balance and faster convergence speed. For Hyperparameters optimization influencing learning rates and success bonus, see Table (\ref{tab:S1}) in Supplementary Note~3.

Here, the actor-critic architecture used for continuous control, includes actor(policy) and twin Q-critics network, each with corresponding target networks. The actor receives a $20$D state as input and outputs a $3$D action distribution via two hidden layers with $256$ ReLU units. Critics process $23$D state--action inputs using identical architectures. Mean reward and episode length for the best agent evolve as in Fig.(\ref{fig:rewardovertime}) and Fig.(\ref{fig:episodeovertime}) in Supplementary Note~3.

\section*{Data availability}
Data generated during the study is available upon reasonable request.

\section*{Code availability}
Codes designed for the study are available upon reasonable request.

\section*{Author contributions}
M.P. proposed the problem. A.G. developed the HA. A.G. and R.O. performed RL simulations. A.S.H. wrote the manuscript with inputs from R.O. The overall project was managed by A.S.H. All the authors contributed to the discussion and analysis. 

\section*{Competing interests}
The authors declare no competing interests. 

\section*{Acknowledgements}
The authors thank Piotr Kolendreski, Andrzej Wodecki and Michał Chromiak for discussions. The current work was a part of the project `Self-calibrating electronic controller for satellite quantum key generator', contract no. P005-SYD-SEKSGSK-001 with AROBS (SYDERAL) Polska. The project was initiated when A.S.H. and M.P. were funded by Foundation for Polish Science (IRAP Project, ICTQT, contract no. MAB/2018/5, co-financed by EU within Smart Growth Operational Programme). M.P. thanks NCN Poland, ChistEra-2023/05/Y/ST2/00005 under the project Modern Device Independent Cryptography (MoDIC).

\appendix
\section*{Supplementary Note 1: Characterization of the entanglement source} \label{supp:entsource}
\subsection{Simulation parameters and methods}
Numerical simulations for modeling spontaneous parametric down-conversion (SPDC) process in a 10~mm periodically poled lithium niobate (PPLN) crystal with periodicity $\Lambda=19.388\mu$m was carried out. A collinear type-0 phase matching was assumed, with pump, signal and idler photons polarized along the extraordinary axis. The pump was a monochromatic 775~nm field propagating along the crystal $z$-axis.

Temperature-dependent phase-matching conditions were evaluated using Sellmeier equations for LiNbO$_3$ incorporating temperature dependence of the refractive indices. Quasi-phase matching (QPM) included grating wavevector $\vec{K}=\frac{2\pi}{\Lambda}$ with poling period $\Lambda$. Simulations were performed in the low-gain regime, neglecting optical losses, pump depletion and higher-order dispersion.

The spectral and spatial properties of the photon pairs were extracted from the phase-matching function. These include the signal and idler wavelengths, emission (opening) angles and the longitudinal wavevector mismatch as functions of temperature. From the spectral distributions, mean wavelength, modal wavelength, standard deviation and full width at half maximum (FWHM) were computed. A relative estimate of the source brightness using overlap between pump field and the effective phase-matching function provides a measure of the photon-pair generation efficiency.

\begin{figure}[ht!]
    \centering
    \includegraphics[width=0.45\textwidth]{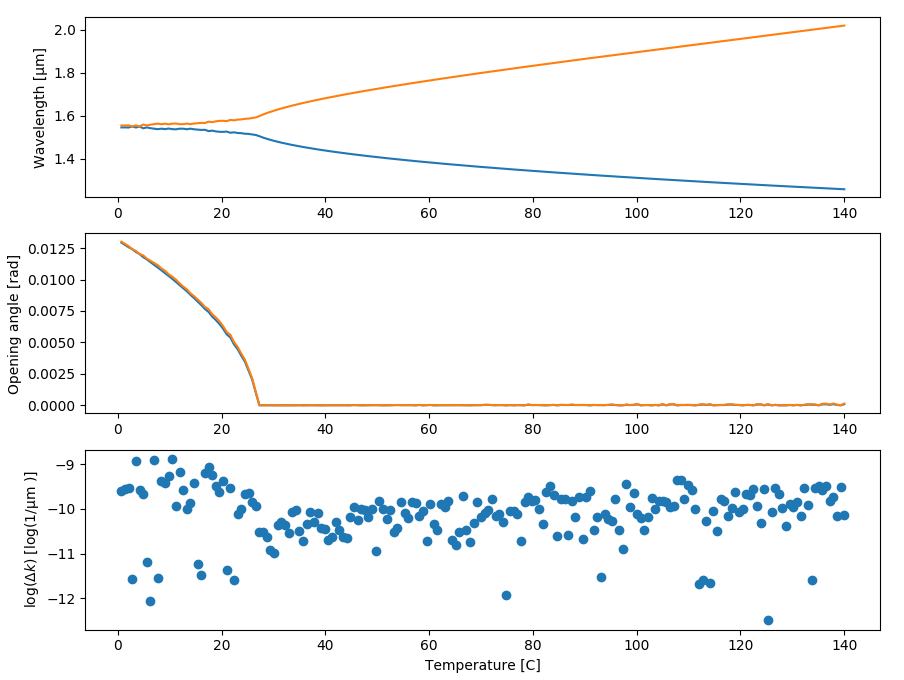}
    \captionsetup{justification=Justified, singlelinecheck=off}
    \caption{The dependence of OPO on temperature is obtained from the spatial and spectral characteristics of the signal and the idler photon pairs. The plots represent the wavelength, opening angle and wavevector mismatch of these photons as a function of temperature respectively. For temperatures above $25^{\circ}C$, the wavelengths of the signal and the idler photons are farther apart and their opening angle is $0$. The logarithm of the wavevector mismatch indicates that efficiency of SPDC at higher temperatures reduces.}
    \label{fig:OPO}
    \end{figure}

\subsection{Temperature dependence and Optical Parametric Oscillator (OPO)}
Figure~\ref{fig:OPO} shows the temperature dependence of OPO-characteristics from the spatial and spectral properties of the signal and idler photons. As the crystal temperature is varied, the QPM shifts, resulting in changes in the wavelengths, emission angles and wavevector mismatch of the photon pairs.

For temperatures below approximately $25^{\circ}$C, the phase-matching condition is not satisfied, leading to a large longitudinal wavevector mismatch and a strong suppression of SPDC. If the temperature exceeds this threshold, efficient QPM is achieved. In this regime, the signal and idler wavelengths differ by several nanometers and the emission becomes increasingly collinear, with the opening angle approaching zero.

The logarithmic dependence of the wavevector mismatch highlights the temperature sensitivity of nonlinear interactions. Small deviations from the optimal value, result in a sharp increase in mismatch, corresponding to a reduction in the conversion efficiency. These results emphasize the importance of precise temperature stabilization for achieving efficient OPO operation and stable photon-pair generation.

\subsection{Spectral properties and source performance}
\begin{figure}[!ht]
    \includegraphics[width=0.51\textwidth]{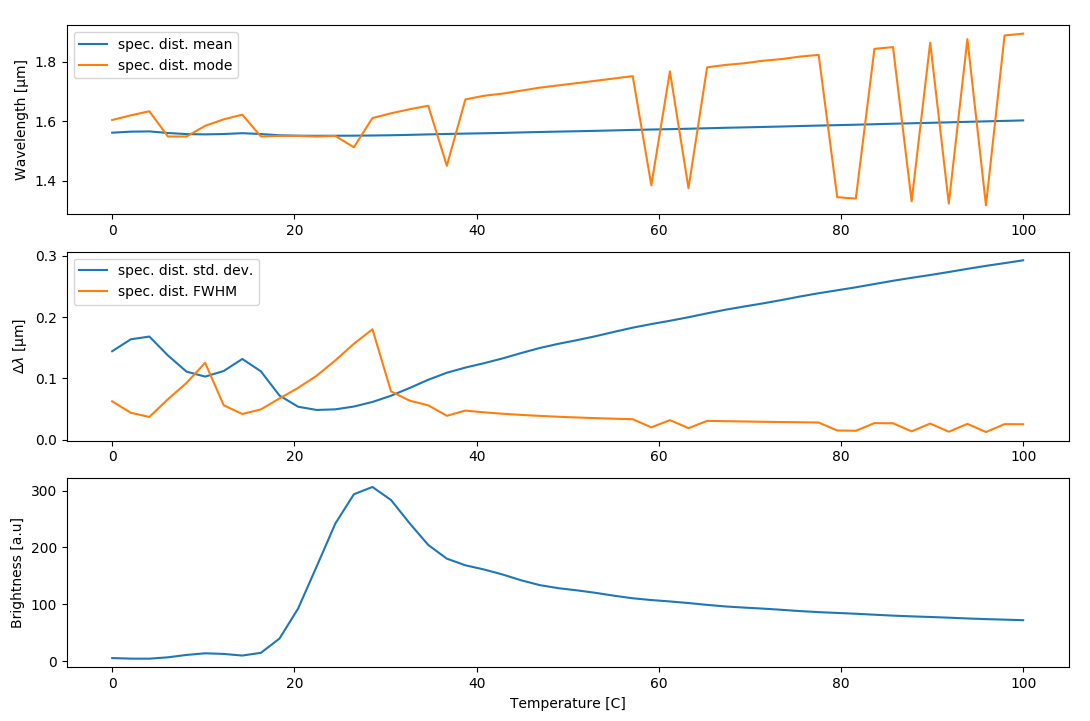}
    \captionsetup{justification=Justified, singlelinecheck=off}
    \caption{The dependence of SPDC on temperature variation is exhibited by the mean, mode, standard deviation and the full width at half maximum of the spectral distribution of the photons. The brightness estimation of the source demonstrates better performance of the source approximately above $25^{\circ}c$. Below it, SPDC vanishes.}
    \label{fig:SPDC}
\end{figure}
The temperature dependence of the spectral properties of SPDC is summarized in Fig.~\ref{fig:SPDC}. The mean and modal wavelengths of the signal and idler shift monotonically with temperature, reflecting the underlying phase-matching dynamics. The spectral bandwidth, characterized by the standard deviation and FWHM, remains narrow within the optimal temperature range and broadens as the phase-matching condition degrades.

The estimated source brightness exhibits a clear maximum for temperatures above $25^{\circ}$C, coinciding with minimal wavevector mismatch and collinear emission. Below this, the brightness rapidly decreases, consistent with vanishing of the SPDC process. These results identify an optimal operating temperature range in which the source has high brightness, stable spectral characteristics and efficient photon-pair generation.

\begin{figure}[!ht]
    \includegraphics[width=0.45\textwidth]{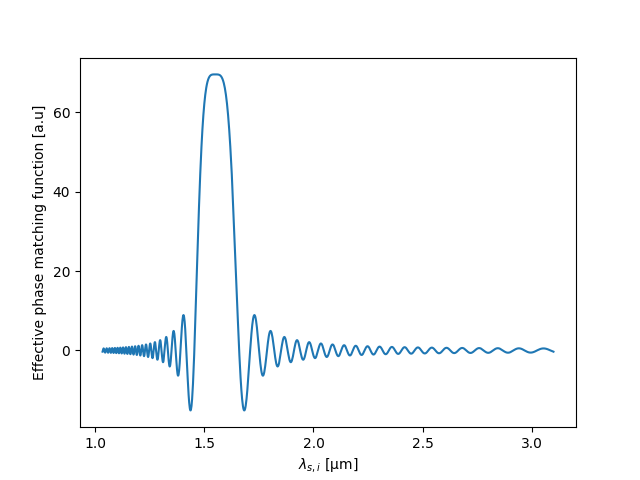}
    \captionsetup{justification=Justified, singlelinecheck=off}
    \caption{Real part of the biphoton wavefunction at room temperature obtained from numerical simulations of the SPDC process. The imaginary part is zero due to the symmetric phase-matching conditions and the absence of pump spectral phase. Because the pump is assumed to be monochromatic, the biphoton state is effectively 1D and its amplitude depends on the effective phase-matching function of the PPLN crystal.}
    \label{fig:Biphoton}
\end{figure}

\subsection{Biphoton wavefunction}
The biphoton wavefunction of the entangled photon pairs at room temperature is shown in Fig.~\ref{fig:Biphoton}. Under the assumption of a monochromatic pump, the wavefunction is effectively one-dimensional, with its amplitude directly proportional to the crystal's phase-matching function. The imaginary part of the amplitude is negligible, reflecting the symmetric phase-matching conditions and the absence of spectral phase in the pump field. 

The shape of the biphoton wavefunction is closely linked to the temperature dependence of the phase-matching condition. As the crystal temperature approaches the optimal quasi-phase-matching point, the wavefunction becomes narrower in the spectral domain corresponding to an increase in the photon-pair generation efficiency and a greater source brightness. Deviations from the optimal operating temperature broadens the wavefunction causing a reduction in its amplitude. This analysis demonstrates that precise temperature control not only affects SPDC efficiency but also determines the spectral correlations and coherence properties of the generated entangled photon pairs.

\section*{Supplementary Note 2: Implementation details of the Heuristic algorithm}
\label{{supp:ha}}
This Supplementary Note provides additional details of the heuristic algorithm (HA) used to optimize the coupling efficiency of the entanglement source. In particular, the flowchart shown in Supplementary Fig.~1 illustrates the sequence of measurements, statistical evaluations and positional adjustments performed during alignment.

\begin{figure}[!hbt]
\centering
\begin{tikzpicture}[node distance=1.75cm, scale=0.65, every node/.style={transform shape}]
    \node (start) [startstop] {Start};
    \node (pro1) [process, below of=start, yshift= 0.2cm] {Measure baseline signal};
    \node (dec1) [decision, below of=pro1, yshift=-0.8cm] {Signal $>$ threshold?};
    \node (abort) [startstop, right of=dec1, xshift=2.3cm] {Abort};
    \node (pro2) [process, below of=dec1,yshift=-0.85cm] {Iterative adjustment along $Z$ and $XY$};
    \node (dec2) [decision, below of=pro2, yshift=-1.2cm] {Max signal or time?};
    \node (stop) [startstop, below of=dec2, yshift=-1.4cm] {Stop};

    \draw [arrow] (start) -- (pro1);
    \draw [arrow] (pro1) -- (dec1);
    \draw [arrow] (dec1.east) -- node[above]{\textcolor{red}{No}} (abort.west);
    \draw[arrow] (abort.north) -- ++(0,0) |- (start);
    \draw [arrow] (dec1.south) -- node[left]{\textcolor{green}{Yes}} (pro2.north);
    \draw [arrow] (pro2.south) -- (dec2.north);
    \draw [arrow] (dec2.south) -- node[left]{\textcolor{green}{Yes}} (stop.north);
    \draw [arrow] (dec2.east) -- ++(1.5,0) |- (pro2.east) node[midway,right]{\textcolor{red}{No}};  
\end{tikzpicture}
\captionsetup{justification=Justified, singlelinecheck=off}
\caption{Flowchart of the HA. The procedure begins with a reference measurement and iterative adjustments along the ($Z$) and ($XY$) positions of the input fiber based on the photon count rate until either a statistically significant signal maximum or a predefined time limit is reached. If the initial signal is below threshold, the algorithm is aborted and restarted.}
\label{alg1}
\end{figure}
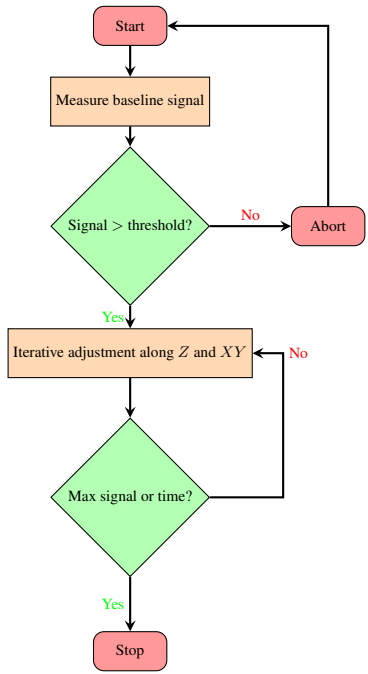

\subsection{Initial positioning and signal verification}
HA begins with a blind jump of $Z_0 = 1580 \mu$m along the $Z$-axis to position the input fiber at the theoretical optimal point, followed by a $120$s integration measurement for maximum sensitivity. This verifies if the detector at the output fiber registers a signal exceeding the background noise at $Z_0$. If not, the algorithm is terminated owing to misalignment beyond automated recovery.

\subsection{Decision metric and hypothesis testing}
After establishing the reference point, HA searches alternating between $Z$ axis and radial $XY$ plane. The $Z$-axis search iteratively adjusts the axial position of the input fiber. If a move leads to a worse result, the step size is reduced and the search direction may be reversed. In the radial plane, the algorithm probes four surrounding positions arranged in a square pattern, identifying the location that yields the highest measured photon count rate. In both these domains, the step sizes are adapted based on the measured signal response and all measurements are evaluated for statistical significance. HA alternates between axial and radial searches until predefined count requirements or time constraints are reached.

The algorithm was tested using a simulated photon source that records photon counts for different misalignment positions. To distinguish true signal improvements from random fluctuations due to photon counting noise, a statistical decision condition is applied. For each measurement performed at a new position, following either a $Z$-axis or radial adjustment, three quantities are recorded: the detected number of counts $N$, the measurement time $S$, and a standardized variable $W$, which serves as the main figure of merit and is defined as
\begin{equation}
W = \frac{\log\!\left(\frac{N}{N_{\mathrm{prev}}}\right)
      - \log\!\left(\frac{S}{S_{\mathrm{prev}}}\right)}
{\sqrt{\frac{1}{N_{\mathrm{prev}}} + \frac{1}{N}}},
\end{equation}
where $N_{\mathrm{prev}}$ and $S_{\mathrm{prev}}$ denote the count and measurement time at the previous position, respectively.

A hypothesis test is performed to determine whether there is no improvement in the mean photon count rate or a measurable increase. Two statistical thresholds corresponding to confidence levels of $99.5\%$ (axial) and $99.9\%$ (radial plane) are defined. If $W$ exceeds one of these thresholds, the null hypothesis is rejected and the new position is accepted as an improvement.

In the radial plane, acceptance further requires that limits on the measurement time and minimum number of counts are satisfied. This prevents the algorithm from spending excessive time in low-signal regions while maintaining fast convergence. For the axial search, two additional thresholds are introduced: a minimum number of counts required to decide that a new position is worse than the previous one and a separate minimum count threshold for confirming improvements. These conditions ensure that the axial search is not dominated by noisy measurements, which is important because axial moves typically produce larger changes in the detected signal and coupling efficiency.

In this search process, the algorithm utilizes a decision tree that analyzes up to four previous measurements categorizing the search status based on consecutive failures. For instance, if a single step degrades the output signal, the algorithm interprets this as an immediate overshoot, triggering a direction reversal and a reduction in step size to $\frac{2}{3}$ of the initial value to precisely re-approach the peak. However, if HA detects a sequence of two or three consecutive measurements that yield worse results, it abandons the current positions and moves the fiber back to the last known "good" coordinate. Thus, resetting the search from a reliable point to prevent further drifts.

The decision outcome directly influences the control flow of HA. If the null hypothesis is rejected, the algorithm continues with further adjustments. If the result is inconclusive or worse than the previous measurement, the algorithm may reduce the step size, reverse the search direction or terminate the current axis search in favor of exploring another. If several consecutive iterations fail to produce a statistically significant improvement, the fiber is repositioned at the most recently verified optimal coordinate, or the search is re-initialized by returning to the predefined optimal $Z$ distance specified in the system parameters. The alignment process stop at $90\%$ theoretical maximum counts or $3600$s. Simulated photon counts across misalignment confirms robust and noise-resistant convergence, preserving rapid alignment.

\section*{Supplementary Note 3: Reinforcement Learning}
\subsection{Hyperparameter optimization}
\label{supp:hyper}

The learning rate and success bonus were identified as the most important hyperparameters affecting training stability and convergence speed. Multiple training runs were performed with different values of these parameters to assess their impact on the learning dynamics. Excessively high learning rates led to unstable training and large reward variance, while lower values slowed convergence. Similarly, insufficient success bonus values resulted in delayed task completion, whereas overly large bonuses encouraged premature convergence to suboptimal behaviors. Supplementary Table~S1 lists the complete set of hyperparameters used for training the Soft Actor--Critic agent. Parameters that were subject to optimization are highlighted in bold.

\begin{table}[!ht]
\centering
\begin{tabular}{|l|l|}
\hline
\textbf{Parameter} & \textbf{Value} \\ \hline
\textbf{Batch size} & 128 \\ \hline
$R_{max}$ & -1 \\ \hline
$R_{min}$ & 1 \\ \hline
Bonus list length & 20 \\ \hline
\textbf{Success bonus} & 0.6 \\ \hline
$t_{step}$ & 31 \\ \hline
$r_{step}^{max}$ & 72 \\ \hline
$z_{step}^{max}$ & 563 \\ \hline
Observation number & 5 \\ \hline
Learning timesteps & 12\,000\,000 \\ \hline
\textbf{Learning rate} & $3 \times 10^{-4}$ \\ \hline
Discount factor ($\gamma$) & 0.0003 \\ \hline
\end{tabular}
\caption{Complete list of hyperparameters used during training. Optimized parameters are shown in bold.}
\label{tab:S1}
\end{table}

\subsection{Training}
\label{supp:train}
\begin{figure}[!ht]
\includegraphics[width=0.5\linewidth]{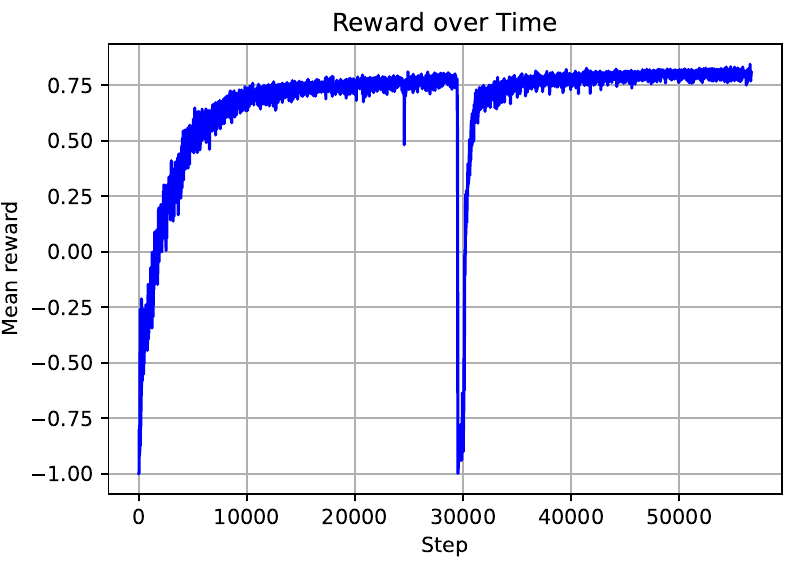}
\caption{Evolution of the mean reward obtained by the best-performing agent during training as a function of the number of environment steps.}
	\label{fig:rewardovertime}
\end{figure}

\begin{figure}[!ht]
	\includegraphics[width=0.5\linewidth]{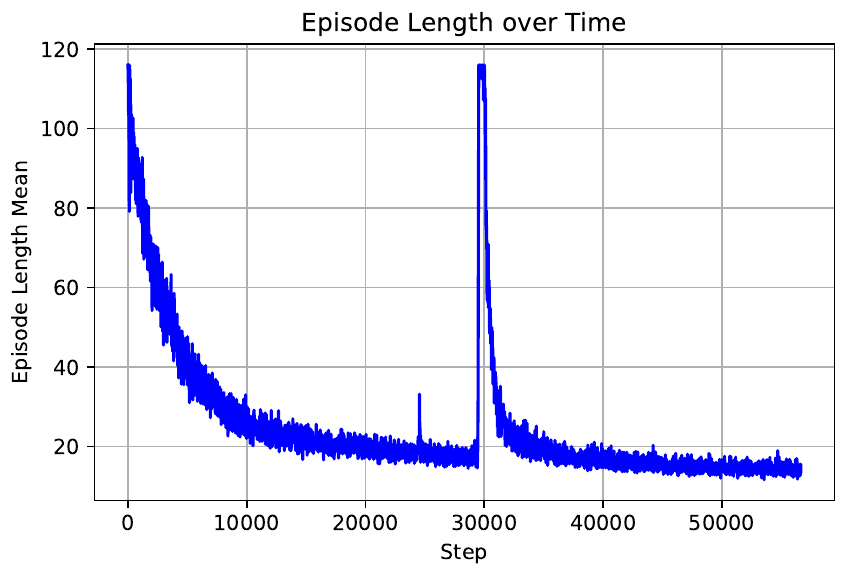}
\caption{Evolution of the episode length obtained by the best-performing agent during training as a function of the number of environment steps.}
	\label{fig:episodeovertime}
\end{figure}

Here, each critic network processes a concatenated state--action input of 23D and uses the same architecture of two hidden layers with 256 units and ReLU activations. Training was performed for a total of 12 million environment interaction steps. The evolution of the mean reward obtained by the best-performing agent during training is shown in Fig.~\ref{fig:rewardovertime}, while the corresponding episode length over training time is presented in Fig.~\ref{fig:episodeovertime}. A transient policy collapse is observed around step $30000$ but a rapid recovery in the learning indicates that the final outcome remains unaffected.

\subsection{Modified ROC-AUC Metric}
ROC curves illustrate the diagnostic ability of binary classifiers by plotting true positive rate (TPR) against false positive rate (FPR) at various thresholds. Thereby aiding the visualization of sensitivity-specificity trade-offs \cite{fawcett2006introduction}. AUC represents the probability that a classifier ranks a random positive instance higher than a random negative one.

Unlike standard ROC, our modification replaces FPR with calibration time $t_{\max_i}$ and cumulative accuracy $A(t_{\max_i})$ (fraction of trials converging within $t_{\max_i}$). The curve gradient represents convergence speed with steeper gradients indicating faster calibration.

The AUC integrates $A(t_{\max_i})$ from $t_{\max_i}=0$ to 60 minutes:
\[
\mathrm{AUC} = \int_0^{60} A(t_{\max_i}) \, dt_{\max_i},
\]
normalized such that AUC = 1.0 when all trials converge at $t=0$, and AUC approaches 0 for no convergence within 60 min. Faster algorithms yield higher AUC due to earlier rise in $A(t)$.

RL reaches near perfect alignment ($A=1.0$) by $\sim$10 min, while HA requires $\sim$30 min, yielding the observed AUC gap. This metric extends naturally to other time-dependent optimization tasks.

\bibliographystyle{h-physrev}
\bibliography{ref.bib}
\end{document}